\documentclass[useAMS,usenatbib]{mn2e}
\usepackage{txfonts}
\usepackage{natbib}
\usepackage[all]{xy}
\usepackage[dvips]{graphicx}

\usepackage{hyperref}
\usepackage[usenames,dvipsnames]{color}
\definecolor{menublue}{rgb}{0.0,0.0,0.5}
\definecolor{citegreen}{rgb}{0.0,1.0,0.0}
\definecolor{urlred}{rgb}{1.0,0.0,0.0}
\hypersetup{bookmarksopen,pdfstartview={FitH},colorlinks=true, breaklinks=true,menucolor=menublue,urlcolor=urlred,citecolor=citegreen,linkcolor=blue}
\usepackage[all]{hypcap}

\usepackage{array}
\newenvironment{mycases}{%
\left\{\array{cc}}{\endarray\right.}

\def\del#1{{}}
\def\spirou#1{{#1}}

\onecolumn

\newcommand{\ltsima}{$\; \buildrel < \over \sim \;$}
\newcommand{\lsim}{\lower.5ex\hbox{\ltsima}}
\newcommand{\gtsima}{$\; \buildrel > \over \sim \;$}
\newcommand{\gsim}{\lower.5ex\hbox{\gtsima}}
\newcommand{\bra}{\langle}
\newcommand{\ket}{\rangle}

\newcommand{\dd}{\mathrm{d}}

\newcommand{\vecx}{\bmath{x}}

\newcommand{\vecl}{\bmath{\ell}}

\newcommand{\trace}{\mathrm{tr}}

\newcommand{\chip}{{\chi^\prime}}

\newcommand{\dirac}{\delta_D}
\newcommand{\heaviside}{\Theta}
\newcommand{\veck}{\bmath{k}}

\newcommand{\fnl}{f_\mathrm{NL}}
\newcommand{\gnl}{g_\mathrm{NL}}
\newcommand{\tnl}{\tau_\mathrm{NL}}

\title[SY-inequality and weak lensing]
{A test of the Suyama-Yamaguchi inequality from weak lensing}
\author[A. Grassi, L. Heisenberg, C.T. Byrnes, B.M. Sch{\"a}fer]{Alessandra Grassi$^1$\thanks{e-mail: sandri@ari.uni-heidelberg.de}, Lavinia Heisenberg$^{2,3}$, Christian T. Byrnes$^4$, Bj{\"o}rn Malte Sch{\"a}fer$^1$\\
$^1$Astronomisches Recheninstitut, Zentrum f{\"u}r Astronomie, Universit{\"a}t Heidelberg, Philosophenweg 12, 69120 Heidelberg, Germany\\
$^2$D\'epartement de Physique Th\'eorique and Center for Astroparticle Physics,
 Universit\'e de Gen\`eve, 24 Quai E. Ansermet, CH-1211,  Gen\`eve, Switzerland\\
$^3$Department of Physics, Case Western Reserve University, 10900 Euclid Ave, Cleveland, OH
44106, United States of America\\
$^4$Department of Physics and Astronomy, University of Sussex, Brighton, BN1 9RH, United Kingdom}

\begin{document}
\pagerange{\pageref{firstpage}--\pageref{lastpage}}
\pubyear{2013}
\maketitle
\label{firstpage}

\begin{abstract}
We investigate the weak lensing signature of primordial non-Gaussianities of the local type by constraining the magnitude of the weak convergence bi- and trispectra expected for the Euclid weak lensing survey. Starting from expressions for the weak convergence spectra, bispectra and trispectra, whose relative magnitudes we investigate as a function of scale, we compute their respective signal to noise ratios by relating the polyspectra's amplitude to their Gaussian covariance using a Monte-Carlo technique for carrying out the configuration space integrations. In computing the Fisher-matrix on the non-Gaussianity parameters $\fnl$, $\gnl$ and $\tnl$ with a very similar technique, we can derive Bayesian evidences for a violation of the Suyama-Yamaguchi relation $\tnl\geq(6\fnl/5)^2$ as a function of the true $\fnl$ and $\tnl$-values and show that the relation can be probed down to levels of $\fnl\simeq10^2$ and $\tnl\simeq10^5$. In a related study, we derive analytical expressions for the probability density that the SY-relation is exactly fulfilled, as required by models in which any one field generates the perturbations. We conclude with an outlook on the levels of non-Gaussianity that can be probed with tomographic lensing surveys.
\end{abstract}


\section{Introduction}
Advances in observational cosmology has made it possible to probe models of the early Universe and the mechanisms that can generate small seed perturbations in the density field from which the cosmic large-scale structure grew by gravitational instability. One of the most prominent of these models is inflation, in which the Universe underwent an extremely rapid exponential expansion and where small fluctuations in the inflationary field gave rise to fluctuations in the gravitational potential and which then imprinted these fluctuations onto all cosmic fluids \citep[for reviews, see][]{2004PhR...402..103B,Seery:2007zr, Komatsu:2009kd, Komatsu:2010qf, 2010AdAst2010E..89D,2010CQGra..27l4011D, 2010AdAst2010E..64V,  2011arXiv1104.0926J, Wang:2013zr, Martin:2013dq, Lesgourgues:2013ij}. Observationally, inflationary models can be distinguished by the spectral index $n_s$ along with a possible scale dependence, the scalar to tensor-ratio $r$ and, perhaps most importantly, the non-Gaussian signatures, quantified by $n$-point correlation functions or by polyspectra of order $n$ in Fourier-space. They are of particular interest as there is a relation between the statistical properties of the fields and its dynamics. Additionally, the configuration space dependence of the polyspectra yields valuable information on the type of inflationary model \citep{Byun:2013nx}.

The (possibly non-Gaussian) density fluctuations are subsequently imprinted in the cosmic microwave background (CMB) as temperature anisotropies \citep{Fergusson:2008ra,2007PhRvD..76h3523F, 2009MNRAS.397..837V, 2010PhRvD..82b3502F, Pettinari:2013hc}, in the matter distribution which can be probed by e.g.~gravitational lensing and in the number density of galaxies. Hereby it is advantageous that the observable is linear in the field whose statistical property we investigate. In case of linear dependence the $n$-point functions of the observable field can be mapped directly onto the corresponding $n$-point function of the primordial density perturbation, which reflects the microphysics of the early Universe.

The first important measurement quantifying non-Gaussianity is the parameter $\fnl$ which describes the skewness of inflationary fluctuations and determines the amplitude of the bispectrum. Not only the bispectrum but also the trispectrum can successfully be constrained by future precisions measurements, where the parameters $\gnl$ and $\tnl$ determine the trispectrum amplitude. The complementary analysis of both the bi- and the trispectra in the future experiments will make us able to extract more information about the mechanism of generating the primordial curvature perturbations and constrain the model of the early Universe. Therefore, it is an indispensable task for cosmology to obtain the configuration space dependence for the higher polyspectra and to make clear predictions for the non-Gaussianity parameters. The non-Gaussianities are commonly expressed as perturbations of modes of the potential $\propto k^{n_s/2-2}$ but can in principle have scale dependences \citep{Chen:2005fe, 2008JCAP...04..014L, Sefusatti:2009xu, Riotto:2010nh,2010JCAP...10..004B,Becker:2010hx,Byrnes:2010xd}.

The first cosmological data release of the Planck satellite has resulted in the tighest ever constraints on $\fnl$ and $\tnl$ \citep{Planck-Collaboration:2013ve}. For the local bispectrum, $\fnl=2.7\pm5.8$, with the $1\sigma$ confidence level quoted, while the 95\% upper bound on the trispectrum parameter is $\tnl\leq 2800$. The $\fnl$ is about a factor of 4 improvement over the WMAP bound \citep{2012arXiv1212.5225B, Giannantonio:2013uqa}, while the $\tnl$ bound is improved by about an order of magnitude \citep{2012MNRAS.425.2187H}. No Planck bound on $\gnl$ has yet been made, the tightest bound is currently $\gnl=(-3.3\pm2.2)\times 10^5$ from WMAP9 data \citep{2013arXiv1303.4626S}. Previous CMB constraints were made in \citep{2008MNRAS.389.1439H, 2010arXiv1001.5026S, 2010arXiv1012.6039F}. The bound on $\fnl$ is close to cosmic variance limited for any CMB experiment, for the trispectrum parameters the bounds may still improve by a factor of a few, see e.g.~\citep{2010arXiv1001.5026S, 2010PhRvD..82b3502F, 2013arXiv1303.4626S}.


An alternative way of constraining non-Gaussianities are the number density of clusters as a function of their mass, see \citet{2011MNRAS.414.1545F, LoVerde:2011kl, Enqvist:2011oq} who show that constraints of order $10^2$ on $\fnl$ and $10^8$ on $\gnl$.

In comparison to other probes, weak gravitational lensing provides weaker bounds, but non-Gaussianities have nevertheless important implications for weak lensing. Although the weak lensing bispectrum is by far dominated by structure formation non-Gaussianities \citep{2003MNRAS.344..857T, 2003A&A...397..405B, 2004MNRAS.348..897T}, whose observational signature has been detected at high significance, \citep[via the quasar magnification bias and the aperture mass skewness, ][respectively]{2003A&A...409..411M, 2011MNRAS.410..143S}, there are a number of studies focusing on primordial non-Gaussianities, for example weak lensing peak counts \citep{2011ApJ...728L..13M}, yielding $\sigma_{\fnl}\simeq 10$ constraints on non-Gaussianities, or topological measures of the weak lensing map, for instance the skeleton \citep{2011arXiv1103.5396F} or Minkowski functionals \citep{2011arXiv1103.1876M}. Direct estimation of the inflationary weak lensing bispectra is possible \citep{2011MNRAS.411..595P, Schafer:2012kl} but suffers from the Gaussianising effect of the line of sight-integration \citep{Jeong:2011fk}. Similar to the weak lensing spectrum, bispectra also suffer from contamination by intrinsic alignments \citep{2008MNRAS.388..991S} and baryonic physics \citep{2011arXiv1105.1075S}.

The description of inflationary non-Gaussianities is done in a perturbative way and for the relative magnitude of non-Gaussianities of different order the Suyama-Yamaguchi (SY) relation applies \citep{Suyama:2008pi, Suyama:2010fu, Lewis:2011au, Smith:2011mi, Sugiyama:2012lh, Assassi:2012zq, Kehagias:2012pd, Beltran-Almeida:2013ff, Rodriguez:2013ys, 2013PhRvD..87d3512T}, which in the most basic form relates the amplitudes of the bi- and of the trispectrum. Recently, it has been proposed that testing for a violation of the SY-inequality would make it possible to distinguish between different classes of inflationary models. In this work we focus on the relation between the non-Gaussianity parameters $\fnl$ and $\tnl$ for a local model, and investigate how well the future Euclid survey can probe the SY-relation: The question we address is how likely would we believe in the SY-inequality with the infered $\fnl$ and $\tnl$-values. We accomplish this by studying the Bayesian evidence \citep{2007MNRAS.378...72T, 2008ConPh..49...71T} providing support for the SY-inequality.

Models in which a single field generates the primordial curvature perturbation predict an equality between one term of the trispectrum and the bispectrum, $\tnl=(6\fnl/5)^2$ \citep[provided that the loop corrections are not anomalously large, if they are then $\gnl$ should also be observable][]{2013PhRvD..87d3512T}. Violation of this consistency relation would prove that more than one light field present during inflation had to contribute towards the primordial curvature perturbation. However a verification of the equality would not imply single field inflation, rather that only one of the fields generated perturbations. In fact any detection of non-Gaussianity of the local form will prove that more than one field was present during inflation, because single field inflation predicts negligible levels of local non-Gaussianity. A detection of $\tnl>(6\fnl/5)^2$ would prove that not only that inflation was of the multi-field variety, but also that multiple-fields contributed towards the primordial perturbations, which are the seeds which gave rise to all the structure in the universe today. \spirou{Weaker forms of the SY-relation, $\tnl>(6/5\fnl)^2/2$, has been proposed by \citet{2011PhRvL.106y1301S} for multifield-inflationary models although these may have been refuted by \citet{2011PhRvL.107s1301S}.}

A violation of the Suyama-Yamaguchi inequality would come as a big surprise, since the inequality has been proved to hold for all models of inflation. Even more strongly, in the limit of an infinite volume survey it holds true simply by the definitions of $\tnl$ and $\fnl$, regardless of the theory relating to the primordial perturbations. However since realistic surveys will always have a finite volume, a breaking of the inequality could occur. It remains unclear how one should interpret a breaking of the inequality, and whether any concrete scenarios can be constructed in which this would occur. A violation may be related to a breaking of statistical homogeneity \citep{Smith:2011mi}.

After a brief summary of cosmology and structure formation in Sect.~\ref{sect_cosmology} we introduce primordial non-Gaussianities in Sect.~\ref{sect_nong} along with the SY-inequality relating the relative non-Gaussianity strengths in the polyspectra of different order. The mapping of non-Gaussianities by weak gravitational lensing is summarised in Sect.~\ref{sect_weak_lensing}. Then, we investigate the attainable signal to noise-ratios (Sect.~\ref{sect_s2n}), address degeneracies in the measurement of $\gnl$ and $\tnl$ in (Sect.~\ref{sect_fisher}), carry out statistical tests of the SY-inequality (Sect.~\ref{sect_sy}), investigate analytical distributions of ratios of non-Gaussianity parameters (Sect.~\ref{sect_ana}) and quantify the Bayesian evidence for a violation of the SY-inequality from a lensing measurement (Sect.~\ref{sect_bayes}). We summarise our main results in Sect.~\ref{sect_summary}.

The reference cosmological model used is a spatially flat $w$CDM cosmology with adiabatic initial perturbations for the cold dark matter. The specific parameter choices are $\Omega_m = 0.25$, $n_s = 1$, $\sigma_8 = 0.8$, $\Omega_b=0.04$. The Hubble parameter is set to $h=0.7$ and the Hubble-distance is given by $c/H_0=2996.9~\mathrm{Mpc}/h$. The dark energy equation of state is assumed to be constant with a value of $w=-0.9$. We prefer to work with these values that differ slightly from the recent Planck results \citep{Planck-Collaboration:2013ly} because lensing prefers lower $\Omega_m$-values and larger $h$-values \citep{2013arXiv1303.1808H}. Scale-invariance for $n_s$ was chosen for simplicity and should not strongly affect the conclusions as the range of angular scales probed is small and close to the normalisation scale.

The fluctuations are taken to be Gaussian perturbed with weak non-Gaussianities of the local type, and for the weak lensing survey we consider the case of Euclid, with a sky coverage of $f_\mathrm{sky}=1/2$, a median redshift of $0.9$, a yield of $\bar{n}=40~\mathrm{galaxies}/\mathrm{arcmin}^2$ and a ellipticity shape noise of $\sigma_\epsilon=0.3$ \citep{2007MNRAS.381.1018A, 2009ExA....23...17R}.

\section{Cosmology and structure formation}\label{sect_cosmology}
In spatially flat dark energy cosmologies with the matter density parameter $\Omega_m$, the Hubble function $H(a)=\dd\ln a/\dd t$ is given by
\begin{equation}
\frac{H^2(a)}{H_0^2} = \frac{\Omega_m}{a^{3}} + \frac{1-\Omega_m}{a^{3(1+w)}},
\end{equation}
for a constant dark energy equation of state-parameter $w$. The comoving distance $\chi$ and scale factor $a$ are related by
\begin{equation}
\chi = c\int_a^1\:\frac{\dd a}{a^2 H(a)},
\end{equation}
given in units of the Hubble distance $\chi_H=c/H_0$. For the linear matter power spectrum $P(k)$ which describes the Gaussian fluctuation properties of the linearly evolving density field $\delta$,
\begin{equation}
\bra \delta(\bmath{k})\delta(\bmath{k}^\prime)\ket = (2\pi)^3\delta_D(\bmath{k}+\bmath{k}^\prime)P(k)
\end{equation}
the ansatz $P(k)\propto k^{n_s} T^2(k)$ is chosen with the transfer function $T(k)$, which is well approximated by the fitting formula
\begin{equation}
T(q) = \frac{\ln(1+2.34q)}{2.34q}\times \left[1+3.89q+(16.1q)^2+(5.46q)^3+(6.71q)^4\right]^{-1/4},
\end{equation}
for low-matter density cosmologies \citep{1986ApJ...304...15B}. The wave vector $k=q\Gamma$ enters rescaled by the shape parameter $\Gamma$ \citep{1995ApJS..100..281S},
\begin{equation}
\Gamma=
\Omega_m h\exp\left[-\Omega_b\left(1+\frac{\sqrt{2h}}{\Omega_m}\right)\right].
\end{equation}
The fluctuation amplitude is normalised to the variance $\sigma_8^2$,
\begin{equation}
\sigma_R^2 = \int\frac{k^2\dd k}{2\pi^2}\: W^2_R(k)\:P(k),
\end{equation}
with a Fourier-transformed spherical top-hat $W_R(k)=3j_1(kR)/(kR)$ as the filter function operating at $R=8~\mathrm{Mpc}/h$. $j_\ell(x)$ denotes the spherical Bessel function of the first kind of order $\ell$ \citep{1972hmf..book.....A}. The linear  growth of the density field, $\delta(\bmath{x},a)=D_+(a)\delta(\bmath{x},a=1)$, is described by the growth function $D_+(a)$, which is the solution to the growth equation \citep{1997PhRvD..56.4439T, 1998ApJ...508..483W, 2003MNRAS.346..573L},
\begin{equation}
\frac{\dd^2}{\dd a^2}D_+(a) + \frac{1}{a}\left(3+\frac{\dd\ln H}{\dd\ln a}\right)\frac{\dd}{\dd a}D_+(a) = 
\frac{3}{2a^2}\Omega_m(a) D_+(a).
\label{eqn_growth}
\end{equation}
From the CDM-spectrum of the density perturbation the spectrum of the Newtonian gravitational potential can be obtained
\begin{equation}
P_\Phi(k) = \left(\frac{3\Omega_m}{2\chi_H^2}\right)^2\:k^{n_s-4}\:T(k)^2
\end{equation}
by application of the Poisson-equation which reads $\Delta\Phi = 3\Omega_m/(2\chi_H^2)\delta$ in comoving coordinates at the current epoch, $a=1$.

\section{non-Gaussianities}\label{sect_nong}
Inflation has been a very successful paradigm for understanding the origin of the perturbations we observe in different observational channels today. It explains in a very sophisticated way how the universe was smoothed during a quasi-de Sitter expansion while allowing quantum fluctuations to grow and become classical on superhorizon scales. In its simplest implementation, inflation generically predicts almost  Gaussian density perturbations close to scale-invariance. In the most basic models of inflation fluctuations originate from a single scalar field in approximate slow roll and deviations from the ideal Gaussian statistics is caused by deviations from the slow-roll conditions. Hence, a detection of non-Gaussianity would be indicative of the shape of the inflaton potential or would imply a more elaborate inflationary model. Although there is consensus that competitive constraints on the non-Gaussianity parameters will emerge from CMB-observations and the next generation of large-scale structure experiments, non-Gaussianities beyond the trispectrum will remain difficult if not impossible to measure. For that reason, we focus on the extraction of bi- and trispectra from lensing data and investigate constraints on their relative magnitude. 

Local non-Gaussianities are described as quadratic and cubic perturbations of the Gaussian potential $\Phi_G(\vecx)$ at a fixed point $\vecx$, which yields in the single-source case the resulting field $\Phi(\vecx)$ \citep{LoVerde:2011kl},
\begin{equation}
\Phi_G(\vecx) \rightarrow \Phi(\vecx) = \Phi_G(\vecx)+ \fnl\left(\Phi_G^2(\vecx) - \bra\Phi_G^2\ket\right) + \gnl\left(\Phi_G^3(\vecx)-3\bra\Phi_G^2\ket\Phi_G(\vecx)\right),
\label{eqn_nong_transformation}
\end{equation}
with the parameters $\fnl$, $\gnl$ and $\tnl$. These perturbations generate in Fourier-space a bispectrum $\bra\Phi(\veck_1)\Phi(\veck_2)\Phi(\veck_3)\ket = (2\pi)^3\dirac(\veck_1+\veck_2+\veck_3)\:B_\Phi(\veck_1,\veck_2,\veck_3)$,
\begin{equation}
B_\Phi(\veck_1,\veck_2,\veck_3) = \left(\frac{3\Omega_m}{2\chi_H^2}\right)^3\:2\fnl\:
\left((k_1k_2)^{n_s-4} + \mathrm{2~perm.}\right)\:
T(k_1)T(k_2)T(k_3),
\end{equation}
and a trispectrum $\bra\Phi(\veck_1)\Phi(\veck_2)\Phi(\veck_3)\Phi(\veck_4)\ket = (2\pi)^3\dirac(\veck_1+\veck_2+\veck_3+\veck_4)\:T_\Phi(\veck_1,\veck_2,\veck_3,\veck_4)$,
\begin{equation}
T_\Phi(\veck_1,\veck_2,\veck_3,\veck_4) = 
\left(\frac{3\Omega_m}{2\chi_H^2}\right)^4\:
\left[
6\gnl\:
\left(
(k_1k_2k_3)^{n_s-4} + \mathrm{3~perm.}
\right)\:
+\frac{25}{9}\tnl\:
\left(
(k_1^{n_s-4}k_3^{n_s-4}\left|\veck_1+\veck_2\right|^{n_s-4} + \mathrm{11~perm.}
\right)
\right]
T(k_1)T(k_2)T(k_3)T(k_4).
\label{T}\end{equation}
The normalisation of each mode $\Phi(\veck)$ is derived from the variance $\sigma_8^2$ of the CDM-spectrum $P(k)$.

Calculating the 4-point function of (\ref{eqn_nong_transformation}) one would find the coefficient $(2\fnl)^2$ instead of the factor $25\tnl/9$ in eqn.~(\ref{T}) \citep[see][]{Byrnes:2006vq}. Since eqn.~(\ref{eqn_nong_transformation}) represents single-source local non-Gaussianity (all of the higher order terms are fully correlated with the linear term), this implies the single-source consistency relation $\tnl=(6\fnl/5)^2$. The factor of $25/9$ in eqn.~(\ref{T}) is due to the conventional definition of $\tnl$ in terms of the curvature perturbation $\zeta$, related by $\zeta=5\Phi/3$. In more general models with multiple fields contributing to $\Phi$, the equality between the two non-linearity parameters is replaced by the Suyama-Yamaguchi inequality $\tnl\geq(6\fnl/5)^2$.

\section{Weak gravitational lensing}\label{sect_weak_lensing}

\subsection{Weak lensing potential and convergence}
Weak gravitational lensing probes the tidal gravitational fields of the cosmic large-scale structure by the distortion of light bundles \citep[for reviews, please refer to][]{2001PhR...340..291B,2010CQGra..27w3001B}. This distortion is measured by the correlated deformation of galaxy ellipticities. The projected lensing potential $\psi$, from which the distortion modes can be obtained by double differentiation, 
\begin{equation}
\psi = 2\int\dd\chi\:W_\psi(\chi)\Phi
\label{eqn_lensing_potential}
\end{equation}
is related to the gravitational potential $\Phi$ by projection with the weighting function $W_\psi(\chi)$,
\begin{equation}
W_\psi(\chi) = \frac{D_+(a)}{a}\frac{G(\chi)}{\chi}.
\end{equation}
Born-type corrections are small for both the spectrum \citep{2010A&A...523A..28K} and the bispectrum \citep{2005PhRvD..72h3001D} compared to the lowest-order calculation. The distribution of the lensed galaxies in redshift is incorporated in the function $G(\chi)$,
\begin{equation}
G(\chi) = \int_\chi^{\chi_H}\dd\chip\: p(\chip) \frac{\dd z}{\dd\chip}\left(1-\frac{\chi}{\chip}\right)
\end{equation}
with $\dd z/\dd\chip = H(\chip) / c$. It is common in the literature to use the parameterisation
\begin{equation}
p(z)\dd z = p_0\left(\frac{z}{z_0}\right)^2\exp\left(-\left(\frac{z}{z_0}\right)^\beta\right)\dd z
\quad\mathrm{with}\quad 
\frac{1}{p_0}=\frac{z_0}{\beta}\Gamma\left(\frac{3}{\beta}\right).
\end{equation}
Because of the linearity of the observables following from eqn.~(\ref{eqn_lensing_potential}) moments of the gravitational potential are mapped onto the same moments of the observable with no mixing taking place. At this point we would like to emphasis that the non-Gaussianity in the weak lensing signal is diluted by the line of sight integration, which, according to the central limit theorem, adds up a large number of non-Gaussian values for the gravitational potential with the consequence that the integrated lensing potential contains weaker non-Gaussianities \citep{Jeong:2011fk}.

\subsection{Convergence polyspectra}
Application of the Limber-equation and repeated substitution of $\kappa = \ell^2\psi/2$ allows the derivation of the convergence spectrum $C_\kappa(\ell)$ from the spectrum $P_\Phi(k)$ of the gravitational potential,
\begin{equation}
C_\kappa(\ell) 
= \ell^4
\int_0^{\chi_H}\frac{\dd\chi}{\chi^2}\:W_\psi^2(\chi)P_\Phi(k),
\end{equation}
of the the convergence bispectrum $B_\kappa(\vecl_1,\vecl_2,\vecl_3)$,
\begin{equation}
B_\kappa(\vecl_1,\vecl_2,\vecl_3) =
(\ell_1\ell_2\ell_3)^2
\int_0^{\chi_H}\frac{\dd\chi}{\chi^4}\:W_\psi^3(\chi)B_\Phi(\veck_1,\veck_2,\veck_3)
\end{equation}
and of the convergence trispectrum $T_\kappa(\vecl_1,\vecl_2,\vecl_3,\vecl_4)$,
\begin{equation}
T_\kappa(\vecl_1,\vecl_2,\vecl_3,\vecl_4)=
(\ell_1\ell_2\ell_3\ell_4)^2
\int_0^{\chi_H}\frac{\dd\chi}{\chi^6}\:W_\psi^4(\chi)T_\Phi(\veck_1,\veck_2,\veck_3,\veck_4).
\end{equation}
This relation follows from the expansion of the tensor $\bpsi = \partial^2\psi/\partial\theta_i\partial\theta_j$ into the basis of all symmetric $2\times2$-matrices provided by the Pauli matrices $\sigma_\alpha$ \citep{1972hmf..book.....A}. In particular, the lensing convergence is given by $\kappa = \trace(\bpsi\sigma_0)/2 = \Delta\psi/2$ with the unit matrix $\sigma_0$. Although the actual observable in lensing are the weak shear components $\gamma_+=\trace(\bpsi\sigma_1)/2$ and $\gamma_\times=\trace(\bpsi\sigma_3)/2$, we present all calculations in terms of the convergence, which has identical statistical properties and being scalar, is easier to work with.

\begin{figure}
\begin{center}
\resizebox{0.5\hsize}{!}{\includegraphics{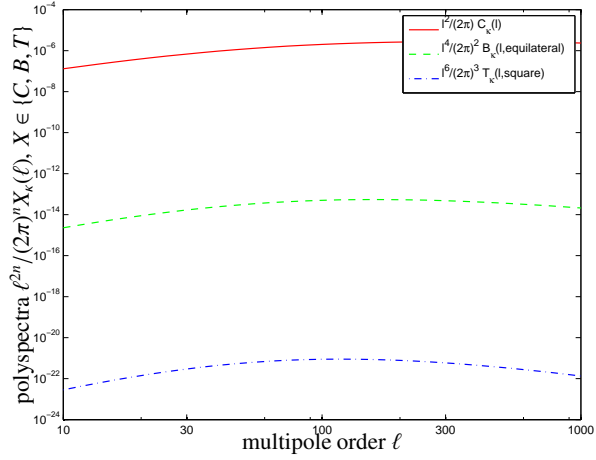}}
\end{center}
\caption{The weak convergence spectrum $C_\kappa(\ell)$ (red solid line), the weak convergence bispectrum for the equilateral configuration $B_\kappa(\ell)$ for an equilateral configuration (green solid line) with $\fnl = 1$, and the convergence trispectrum $T_\kappa(\ell)$ for a square configuration as a function of multipole order $\ell$, for $\gnl=1$ (blue dashed line).}
\label{fig_polyspectra}
\end{figure}

Fig.~\ref{fig_polyspectra} shows the weak lensing spectrum and the non-Gaussian bi- and trispectra as a function of multipole order $\ell$. For the bispectrum we choose an equilateral configuration and for the trispectrum a square one, which are in fact lower bounds on the bi- and trispectrum amplitudes for local non-Gaussianities. The polyspectra are multiplied with factors of $(\ell)^{2n}$ for making them dimensionless and in that way we were able to show all spectra in a single plot, providing a better physical interpretation of variance, skewness and kurtosis per logarithmic $\ell$-interval. In our derivation we derive the lensing potential directly from the gravitational potential, in which the polyspectra are expressed and subsequently apply $\ell^2$-prefactors to obtain the polyspectra in terms of the weak lensing convergence, for which the covariance and the noise of the measurement is most conveniently expressed. The disadvantage of this method is that the $\tnl$-part of the trispectrum $T_\psi$ diverges for the square configuration, because opposite sides of the square cancel in the $\left|\veck_i-\veck_{i+2}\right|$-terms which can not be exponentiated with a negative number $n_s-4$. We control this by never letting the cosine of the angle between $\veck_i$ and $\veck_{i+2}$ drop below $-0.95$. We verified that this exclusion cone of size $\simeq 20^\circ$ has a minor influence on the computation of signal to noise-ratios.

\begin{figure}
\begin{center}
\resizebox{0.5\hsize}{!}{\includegraphics{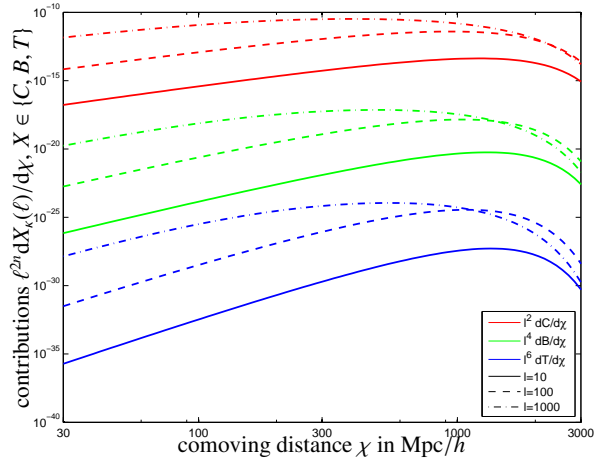}}
\end{center}
\caption{Contributions $\dd C_\kappa(\ell)/\dd\chi$ (red lines), $\dd B_\kappa(\ell)/\dd\chi$ (green lines) for the equilateral configuration and $\dd T_\kappa(\ell)/\dd\chi$ (blue lines) for the square configuration, as a function of comoving distance $\chi$. The non-Gaussianity parameters are chosen to be $\fnl=1$ and $\gnl = 1$. We compare the contributions at $\ell=10$ (solid line) with $\ell=100$ (dashed line) and $\ell=1000$ (dash-dotted line).}
\label{fig_contribution}
\end{figure}

The contributions to the weak lensing polyspectra as a function of comoving distance $\chi$ are shown in Fig.~\ref{fig_contribution}, which is the derivative of Fig.~\ref{fig_polyspectra} at fixed $\ell$. At the same time, the plot presents the integrand of the Limber equation and it demonstrates nicely that the largest contribution to the weak lensing polyspectra comes from the peak of the galaxy distribution, with small variations with multipole order as higher multipoles acquire contributions from slightly lower distances. 

\subsection{Relative magnitudes of weak lensing polyspectra}
The strength of the non-Gaussianity introduced by nonzero values of $\gnl$ and $\tnl$ can be quantified by taking ratios of the three polyspectra. We define the skewness parameter $S(\ell)$ as the ratio
\begin{equation}
S(\ell) = \frac{B_\kappa(\ell)}{C_\kappa(\ell)^{3/2}}
\end{equation}
between the convergence bispectra for the equilateral configuration and the convergence spectrum. In analogy, we define the kurtosis parameter $K(\ell)$,
\begin{equation}
K(\ell) = \frac{T_\kappa(\ell)}{C_\kappa(\ell)^2},
\end{equation}
as the ratio between the convergence trispectrum for the square configuration and the spectrum as a way of quantifying the size of the non-Gaussianity. The relative magnitude of the bi- and trispectrum is given by the function $Q(\ell)$,
\begin{equation}
Q(\ell) = \frac{T_\kappa(\ell)}{B_\kappa(\ell)^{4/3}}.
\end{equation}
For computing the three parameters we set the non-Gaussianity parameters to $\fnl=\gnl=1$.

\begin{figure}
\begin{center}
\resizebox{0.5\hsize}{!}{\includegraphics{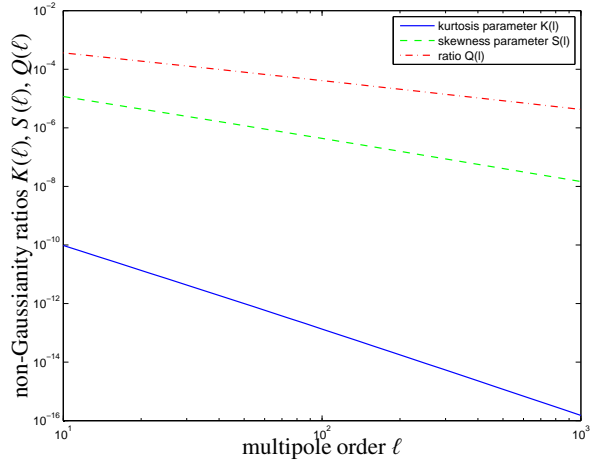}}
\end{center}
\caption{Parameters $K(\ell)$ (blue solid line), $S(\ell)$ (green dashed line) and $Q(\ell)$ (red dash-dotted line), where we chose a equilateral configuration for the convergence bispectrum and a square configuration for the trispectrum. The non-Gaussianity parameters are $\fnl=1$ and $\gnl=1$}
\label{fig_kurtosis}
\end{figure}

The parameters are shown in Fig.~\ref{fig_kurtosis} as a function of multipole order $\ell$. They have been constructed such that the transfer function $T(k)$ in each of the polyspectra is cancelled. The parameters are power-laws because the inflationary part of the spectrum $k^{n_s-4}$ is scale-free and the Wick theorem reduces the polyspectra to products of that inflationary spectrum. The amplitude of the parameters reflects the proportionality of the polyspectra to $3\Omega_m/(2\chi_H^2)$ and the normalisation of each mode proportional to $\sigma_8$. A noticeable outcome in the plot is the fact that the ratio is largest on large scales as anticipated, because the fluctuations in the inflationary fields give rise to fluctuations in the gravitational potential on which the perturbation theory is built. Since the effect of the potential is on large scale and the trispectrum is proportional to the spectrum taken to the third power, the ratio $K(\ell)$ should be the largest on large scales. Therefore as one can see in the Fig.~\ref{fig_kurtosis} the ratio  drops to very small numbers on small scales. Similar arguments apply to $Q(\ell)$ and $S(\ell)$, although the dependences are weaker.

\section{Signal to noise-ratios}\label{sect_s2n}
The signal strength at which a given polyspectrum can be measured is computed as the ratio between that particular polyspectrum and the variance of its estimator averaged over a Gaussian ensemble \citep[which, in the case of structure formation non-Gaussianities, has been shown to be a serious limitation][]{Takada:2009bh, Sato:2013tg, Kayo:2013kx}. We work in the flat-sky approximation because the treatment of the bi- and trispectra involves a configuration-space average, which requires the evaluation of Wigner-symbols in multipole space.

In the flat-sky approximation the signal to noise ratio $\Sigma_C$ of the weak convergence spectrum $C_\kappa(\ell)$ reads \citep{1997ApJ...480...22T, Cooray:2001ve}
\begin{equation}
\Sigma_C^2 = \int\frac{\dd^2\ell}{(2\pi)^2}\:\frac{C_\kappa(\ell)^2}{\mathrm{cov}_C(\ell)},
\end{equation}
with the Gaussian expression for the covariance $\mathrm{cov}_C(\ell)$ \citep{2001ApJ...554...67H, Takada:2013cr},
\begin{equation}
\mathrm{cov}_C(\ell) = 
\frac{2}{f_\mathrm{sky}}\frac{1}{2\pi}\:\tilde{C}_\kappa(\ell)^2.
\end{equation}
Likewise, the signal to noise ratio $\Sigma_B$ of the bispectrum $B_\kappa(\ell)$ is given by \citep{Hu:2000dz, 2004MNRAS.348..897T, Babich:2005fk, Joachimi:2009ly}
\begin{equation}
\Sigma_B^2 = 
\int\frac{\dd^2\ell_1}{(2\pi)^2}\:
\int\frac{\dd^2\ell_2}{(2\pi)^2}\:
\int\frac{\dd^2\ell_3}{(2\pi)^2}\:
\frac{B_\kappa^2(\vecl_1,\vecl_2,\vecl_3)}{\mathrm{cov}_B(\ell_1,\ell_2,\ell_3)}
\end{equation}
where the covariance $\mathrm{cov}_B(\ell_1,\ell_2,\ell_3)$ follows from
\begin{equation}
\mathrm{cov}_B(\ell_1,\ell_2,\ell_3) = 
\frac{6\pi}{f_\mathrm{sky}}\frac{1}{(2\pi)^3}
\:\tilde{C}_\kappa(\ell_1)\tilde{C}_\kappa(\ell_2)\tilde{C}_\kappa(\ell_2).
\end{equation}
Finally, the signal to noise ratio $\Sigma_T$ of the convergence trispectrum $T_\kappa$ results from \citep{Zaldarriaga:2000uq, Hu:2001fv, Kamionkowski:2011vn}
\begin{equation}
\Sigma_T^2 = 
\int\frac{\dd^2\ell_1}{(2\pi)^2}\:
\int\frac{\dd^2\ell_2}{(2\pi)^2}\:
\int\frac{\dd^2\ell_3}{(2\pi)^2}\:
\int\frac{\dd^2\ell_4}{(2\pi)^2}\:
\frac{T_\kappa^2(\vecl_1,\vecl_2,\vecl_3,\vecl_4)}{\mathrm{cov}_T(\ell_1,\ell_2,\ell_3,\ell_4)},
\end{equation}
with the expression
\begin{equation}
\mathrm{cov}_T(\ell_1,\ell_2,\ell_3,\ell_4) = 
\frac{24\pi}{f_\mathrm{sky}}\frac{1}{(2\pi)^4}
\tilde{C}_\kappa(\ell_1)\tilde{C}_\kappa(\ell_2)\tilde{C}_\kappa(\ell_3)\tilde{C}_\kappa(\ell_4)
\end{equation}
for the trispectrum covariance $\mathrm{cov}_T(\ell_1,\ell_2,\ell_3,\ell_4)$. In all covariances, the fluctuations of the weak lensing signal and the noise are taken to be Gaussian and are therefore described by the noisy convergence spectrum $\tilde{C}_\kappa(\ell)$,
\begin{equation}
\tilde{C}_\kappa(\ell) = C_\kappa(\ell) + \frac{\sigma_\epsilon^2}{\bar{n}},
\end{equation}
with the number of galaxies per steradian $\bar{n}$ and the ellipticity noise $\sigma_\epsilon$.

The configuration space integrations for estimating the signal to noise ratios as well as for computing Fisher-matrices are carried out in polar coordinates with a Monte-Carlo integration scheme \citep[specifically, with the CUBA-library by][who provides a range of adaptive Monte-Carlo integration algorithms]{Hahn:2005uq}. We obtained the best results with the SUAVE-algorithm that uses importance sampling for estimating the values of the integrals.

\begin{figure}
\begin{center}
\resizebox{0.5\hsize}{!}{\includegraphics{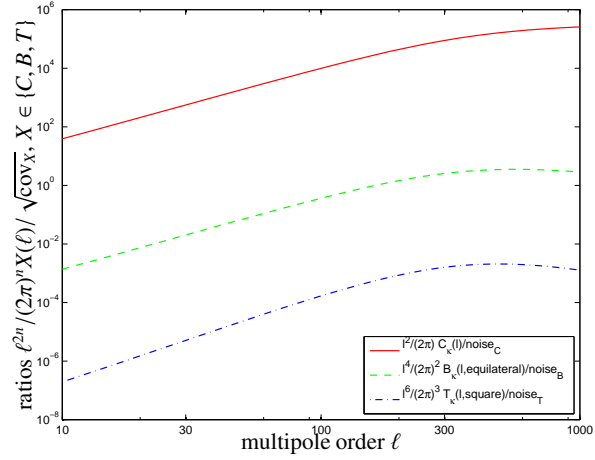}}
\end{center}
\caption{Noise-weighted weak lensing polyspectra: $C_\kappa(\ell)/\sqrt{\mathrm{cov}_C}$ (red solid line), $B_\kappa(\ell)/\sqrt{\mathrm{cov}_B}$ for the equilateral configuration (green dashed line) and $T_\kappa(\ell)/\sqrt{\mathrm{cov}_T}$ (blue dash-dotted line) for the square configuration. The non-Gaussianity parameters are $\fnl=1$ and $\gnl=1$}
\label{fig_noise}
\end{figure}

Fig.~\ref{fig_noise} provides a plot of the polyspectra in units of the noise of their respective estimators. Clearly, the measurements are dominated by cosmic variance and show the according Poissonian dependence with multipole $\ell$, before the galaxy shape noise limits the measurement on small scales and the curves level off or, in the case of the higher polyspectra, begin to drop on multipoles $\ell\gsim300$.

\begin{figure}
\begin{center} 
\resizebox{0.5\hsize}{!}{\includegraphics{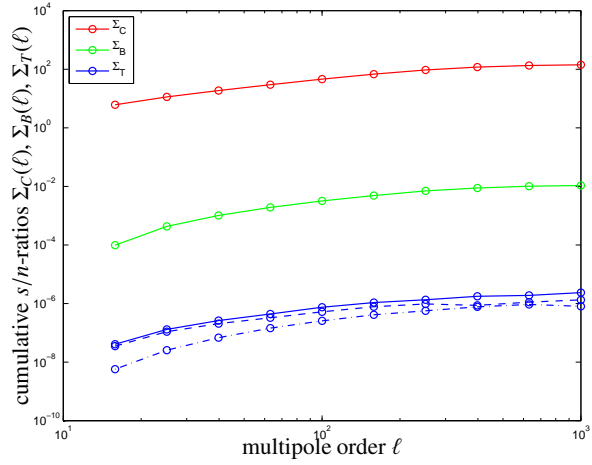}}
\end{center}
\caption{Cumulative signal to noise-ratios $\Sigma_C$ for the weak lensing spectrum (red solid line), $\Sigma_B$ for the weak lensing bispectrum (green solid line) and $\Sigma_T$ for the weak lensing trispectrum, for $(\tnl,\gnl) = (1,1)$ (blue solid line), $(\tnl,\gnl) = (1,0)$ (blue dashed line) and $(\tnl,\gnl) = (0,1)$ (blue dash-dotted line).}
\label{fig_s2n}
\end{figure}

An observation of the polyspectra $C_\kappa(\ell)$, $B_\kappa$ and $T_\kappa$ with Euclid would yield signal to noise ratios as depicted in Fig.~\ref{fig_s2n}. Whereas the convergence spectrum $C_\kappa(\ell)$ can be detected with high significance in integrating over the multipole range up to $\ell=10^3$, the bispectrum would require $\fnl$ to be of the order $10^2$ and the two trispectrum non-Gaussianities $\gnl$ and $\tnl$-values of the order $10^6$ for yielding a detection, which of course is weaker compared to CMB bounds or bounds on the parameters from large-scale structure observation. The reason lies in the non-Gaussianity supression due to the central-limit theorem in the line of sight-integration \citep{Jeong:2011fk}. This could in principle be compensated by resorting to tomographic weak lensing (see Sect.~\ref{sect_summary}).

\section{Degeneracies in the trispectrum}\label{sect_fisher}
\begin{figure}
\begin{center}
\resizebox{0.5\hsize}{!}{\includegraphics{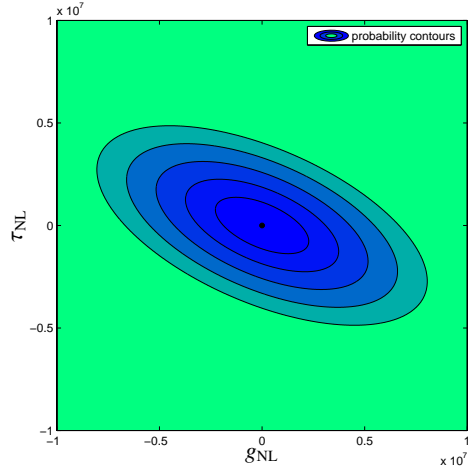}}
\end{center}
\caption{Degeneracies in the $\gnl$-$\tnl$-plane for a measurement with Euclid: The $1\sigma\ldots4\sigma$-contours of the joint likelihood are drawn, while all cosmological parameters are assumed to be known exactly.}
\label{fig_fisher}
\end{figure}

The independency of estimates of $\gnl$ and $\tnl$ from the weak lensing trispectrum are depicted in Fig.~\ref{fig_fisher} where we plot the likelihood contours in the $\gnl$-$\tnl$-plane. The likelihood $\mathcal{L}(\fnl,\gnl,\tnl)$ is taken to be Gaussian,
\begin{equation}
\mathcal{L}(\fnl,\gnl,\tnl) = \sqrt{\frac{\det(F)}{(2\pi)^3}}
\exp\left[
-\frac{1}{2}
\left(
\begin{array}{c}
\fnl \\
\gnl \\
\tnl
\end{array}
\right)^t
\:
F
\:
\left(
\begin{array}{c}
\fnl \\
\gnl \\
\tnl
\end{array}
\right)
\right]
\end{equation}
which can be expected due to the linearity of the polyspectra with the non-Gaussianity parameters. The Fisher-matrix $F$ has been estimated for a purely Gaussian reference model and with a Gaussian covariance, and its entries can be computed in analogy to the signal to noise ratios. The diagonal of the Fisher matrix is composed from the values $\Sigma_B$ and $\Sigma_T$ with the non-Gaussianity parameters set to unity, and the only off-diagonal elements are the two entries $F_{\gnl\tnl}$,
\begin{equation}
F_{\gnl\tnl} = 
\int\frac{\dd^2\ell_1}{(2\pi)^2}\:
\int\frac{\dd^2\ell_2}{(2\pi)^2}\:
\int\frac{\dd^2\ell_3}{(2\pi)^2}\:
\int\frac{\dd^2\ell_4}{(2\pi)^2}\:
\frac{1}{\mathrm{cov}_T(\ell_1,\ell_2,\ell_3,\ell_4)}
T_\kappa(\gnl=1,\tnl=0) T_\kappa(\gnl=0,\tnl=1),
\end{equation}
which again is solved by Monte-Carlo integration in polar coordinates. Essentially, the diagonal elements of the Fisher matrix are given by the inverse squared signal to noise ratios since $B_\kappa\propto\fnl$ and $T_\kappa\propto\tnl$. For Gaussian covariances, the statistical errors on $\fnl$ on one side and $\gnl$ and $\tnl$ on the other are independent, since $F_{\fnl \tnl}=0=F_{\fnl \gnl}$. Clearly, there is a degeneracy that $\gnl$ can be increased at the expense of $\tnl$ and vice versa. In the remainder of the paper, we carry out a marginalisation of the Fisher-matrix such that the uncertainty in $\gnl$ is contained in $\tnl$. The overall precision that can be reached with lensing is about an order of magnitude worse compared to the CMB \citep{Smidt:2010oq}, with a very similar orientation of the degeneracy.

\spirou{We compute the Fisher-matrix on the non-Gaussianity parameters with all other cosmological parameter assumed to a level of accuracy much better than that of $\fnl$, $\gnl$ and $\tnl$, which is reasonable given the high precision one can reach with in particular tomographic weak lensing spectra, baryon acoustic oscillations and the cosmic microwave background. Typical uncertainties are at least two orders of magnitude better than the constraints on non-Gaussianity from weak lensing.}

\section{Testing the Suyama-Yamaguchi-inequality}\label{sect_sy}
Given the fact that there are a vast array of different inflationary models generating local-type non-Gaussianity, it is indispensable to have a classification of these different models into some categories. This can be for instance achieved by using consistency relations among the non-Gaussianity parameters as the SY-relation. In the literature one distinguishes between three main categories of models, the single-source model, the multi-source model and constrained multi-source model. As the name already reveals the single-source model is a model of one field causing the non-linearities. The important representatives of this category include the pure curvaton and the pure modulated reheating scenarios. It is also possible that multiple sources are simultaneously responsible for the origin of density fluctuations. It could be for instance that both the inflaton and the curvaton fields are generating the non-linearities we observe today. In the case of multi-source models the relations between the non-linearity parameters are different from those for the single-source models. Finally, the constrained multi-source models are models in which the loop contributions in the expressions for the power spectrum and non-linearity parameters are not neglected. The classification into these three categories was based on the relation between $\fnl$ and $\tnl$ \citep{2010JCAP...12..030S}. Nevertheless, this will not be enough to discriminate between the models of each category. For this purpose, we will need further relations between $\fnl$ and $\gnl$. Hereby, the models are distinguished by rather if $\gnl$ is proportional to $\fnl$ ($\fnl\sim\gnl$) or enhanced or suppressed compared to $\fnl$. Summarizing, the $\fnl$-$\tnl$ and $\fnl$-$\gnl$ relations will be powerful tools to discriminate models well. In this work we are focusing on the SY-relation between $\fnl$ and $\tnl$.
The Bayesian evidence \citep[for reviews, see][]{2007MNRAS.378...72T, 2008ConPh..49...71T} for the SY-relation $\tnl\geq(6\fnl/5)^2$ can be expressed as the fraction $\alpha$ of the likelihood $\mathcal{L}$ that provides support:
\begin{equation}
\alpha = 
\int\limits_{\tnl\geq(6\fnl/5)^2}\dd\tnl^\prime\:\int\dd\fnl^\prime\:
\mathcal{L}\left(\fnl-\fnl^\prime,\tnl-\tnl^\prime\right).
\end{equation}
Hence $\alpha$ answers the question as to how likely one would believe in the SY-inequality with inferred $\fnl^\prime$ and $\tnl^\prime$-values if the true values are given by $\fnl$ and $\tnl$. Technically, $\alpha$ corresponds to the integral over the likelihood in the $\fnl$-$\tnl$-plane over the allowed region. If $\alpha=1$, we would fully believe in the SY-inequality, if $\alpha=0$ we would think that the SY-relation is violated. Correspondingly, $1-\alpha$ would provide a quantification of the violation of the SY-relation,
\begin{equation}
1-\alpha =
\int\limits_{\tnl<(6\fnl/5)^2}\dd\tnl^\prime\:\int\dd\fnl^\prime\:
\mathcal{L}\left(\fnl-\fnl^\prime,\tnl-\tnl^\prime\right).
\end{equation}

We can formulate the integration over the allowed region as well as an integration over the full $\fnl$-$\tnl$-range of the likelihood multiplied with the Heaviside-function,
\begin{equation}
\alpha = \int\dd\tnl^\prime\:\int\dd\fnl^\prime\:
\mathcal{L}(\fnl-\fnl^\prime,\tnl-\tnl^\prime)\:\heaviside(\tnl-(6/5\fnl)^2).
\end{equation}
This function would play the role of a theoretical prior in the $\fnl$-$\tnl$-plane. In this interpretation, $\alpha$ corresponds to the Bayesian evidence, that means the degree of belief that the SY-inequality is correct. 

\begin{figure}
\begin{center}
\resizebox{0.5\hsize}{!}{\includegraphics{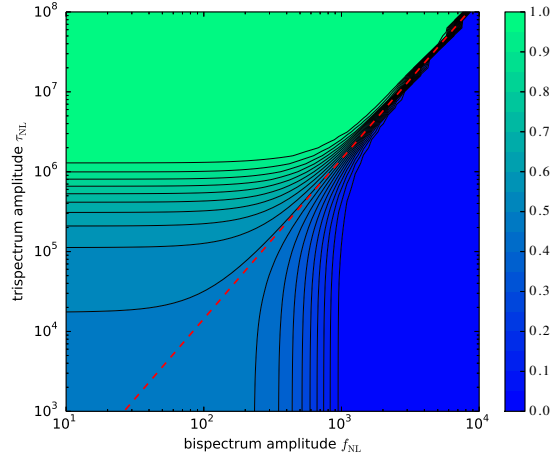}}
\end{center}
\caption{Bayesian evidence $\alpha(\fnl,\tnl)$ in the $\fnl$-$\tnl$-plane. Blue regions correspond to low, green regions to high degrees of belief. The SY-relation $\tnl=(6\fnl/6)^2$ is indicated by the red dashed line.}
\label{fig_sy_fnltnl}
\end{figure}

We can test the SY-inequality $\tnl\geq(6\fnl/5)^2$ up to the errors on $\fnl$ and $\tnl$ provided by the lensing measurement: Fig.~\ref{fig_sy_fnltnl} shows the test statistic $\alpha(\fnl,\gnl,\tnl)$ in the $\fnl$-$\tnl$-plane, where the likelihood has been marginalised over the parameter $\gnl$. The blue regime $\fnl\gsim10^2$ is the parameter space which would not fulfill the SY-inequality, whereas the green area $\tnl\gsim10^5$ is the parameter space where the SY-relation would be fulfilled. Values of $\fnl\lsim10^2$ and $\tnl\lsim10^5$ are inconclusive and even though non-Gaussianity parameters may be inferred that would be in violation of the SY-relation, the wide likelihood would not allow to derive a statement. Another nice feature is the fact that for large $\fnl$ and $\tnl$ the relation can be probed to larger precision and the contours are more closely spaced.

In models where the field which generates non-Gaussianity has a quadratic potential, the non-Gaussianity is mainly captured by $\fnl$, while $\gnl$ is negligible. An example is the curvaton scenario, it is only through self-interactions of the curvaton that $\gnl$ may become large \cite{Enqvist:2009ww}.

\section{analytical distributions}\label{sect_ana}
In this section we derive the analytical expression for the probability density that the SY-relation is exactly fulfilled, $\tnl=(6\fnl/5)^2$, i.e. for the case $(6\fnl/5)^2/\tnl\equiv1$. For this purpose we explore the properties of the distribution 
\begin{equation}
p(Q)\dd Q\quad \mbox{with}\quad Q=\frac{(6\fnl/5)^2}{\tnl}
\end{equation}
where the parameters $\fnl$ and $\tnl$ are both Gaussian distributed with means $\bar \fnl$, $\bar\tnl$ and widths $\sigma_{\fnl}$ and $\sigma_{\tnl}$.

We will split the derivation into two parts. First of all we will derive the distribution for  the product $\fnl^2$. For this purpose we use the transformation of the probability density:
\begin{equation}
p_y(y)\dd y=p_x(x)\dd x
\end{equation}
with the Jacobian $\dd x/\dd y = 1/(2\sqrt{y})$ and where $x=\fnl$ and $y=x^2$. Thus we can write the above equality as
\begin{equation}
p_y(y)=\frac{p_x(\sqrt{y})}{2\sqrt{y}}
\end{equation}
where the probability distribution $p_x(x)$ is given by
\begin{equation}
p_x(\sqrt{y})=\frac{1}{\sqrt{2\pi\sigma_{\fnl}^2}}
\exp\left(-\frac{(\sqrt{y}-\bar \fnl)^2}{2\sigma^2_{\fnl}}\right).
\end{equation}
Naively written in this way, we would lose half of the distribution and do not obtain the right normalization. Therefore we have to distinguish between the different signs of $y$. The distribution of a square of a Gaussian distributed variate $\fnl$ with mean $\bar \fnl$ and variance $\sigma_{\fnl}$ is given by
\begin{eqnarray}\label{eq:distrat}
p_y(y) = 
\frac{1}{\sqrt{2\pi\sigma_{\fnl}^2}}\frac{1}{2\sqrt{y}}\times
\begin{mycases}
\exp\left(-\frac{(\sqrt{y}-\bar \fnl)^2}{2\sigma^2_{\fnl}}\right), & {\mathrm{positive~branch~of~} \sqrt{y}} \\
\exp\left(-\frac{(-\sqrt{-y}-\bar \fnl)^2}{2\sigma^2_{\fnl}}\right), & {\mathrm{negative~branch~of~} \sqrt{y}}
\end{mycases}
\end{eqnarray}
with $y=\fnl^2$. In the special case of normally distributed variates, the above expression would reduce to
\begin{equation}
p_y(y) = 
\frac{1}{\pi\sigma_{\fnl}\sigma_{\tnl}}K_0\left(\frac{|y|}{\sigma_{\fnl}\sigma_{\tnl}}\right)
\end{equation}
where $K_n(y)$ is a modified Bessel function of the second kind \citep{1972hmf..book.....A}. 

The next step is now to implement the distribution eqn.~(\ref{eq:distrat}) into a ratio distribution since we are interested in the distribution of $(6\fnl/5)^2/\tnl$ incorporating the additional factor. The ratio distribution can be written down using the Mellin transformation \citep{2005mmp..book.....A}:
\begin{eqnarray}
p(Q) = \int |\alpha|\dd\alpha\:p_y(\alpha Q,\bar \fnl)p_z(\alpha,\bar \tnl),
\end{eqnarray}
with a Gaussian distribution for $z=\tnl$,
\begin{equation}
p_z(z)=\frac{1}{\sqrt{2\pi\sigma_z^2}}\exp\left(-\frac{(z-\bar z)^2}{2\sigma_z^2}\right).
\end{equation}
In the special case of Gaussian distributed variates with zero mean the distribution would be simply given by the Cauchy distribution \citep{Marsaglia:1965, Marsaglia:2006:JSSOBK:v16i04}, but in the general case eqn.~(\ref{eq:distrat}) needs to be evaluated analytically.

\begin{figure}
\begin{center}
\resizebox{0.7\hsize}{!}{\includegraphics{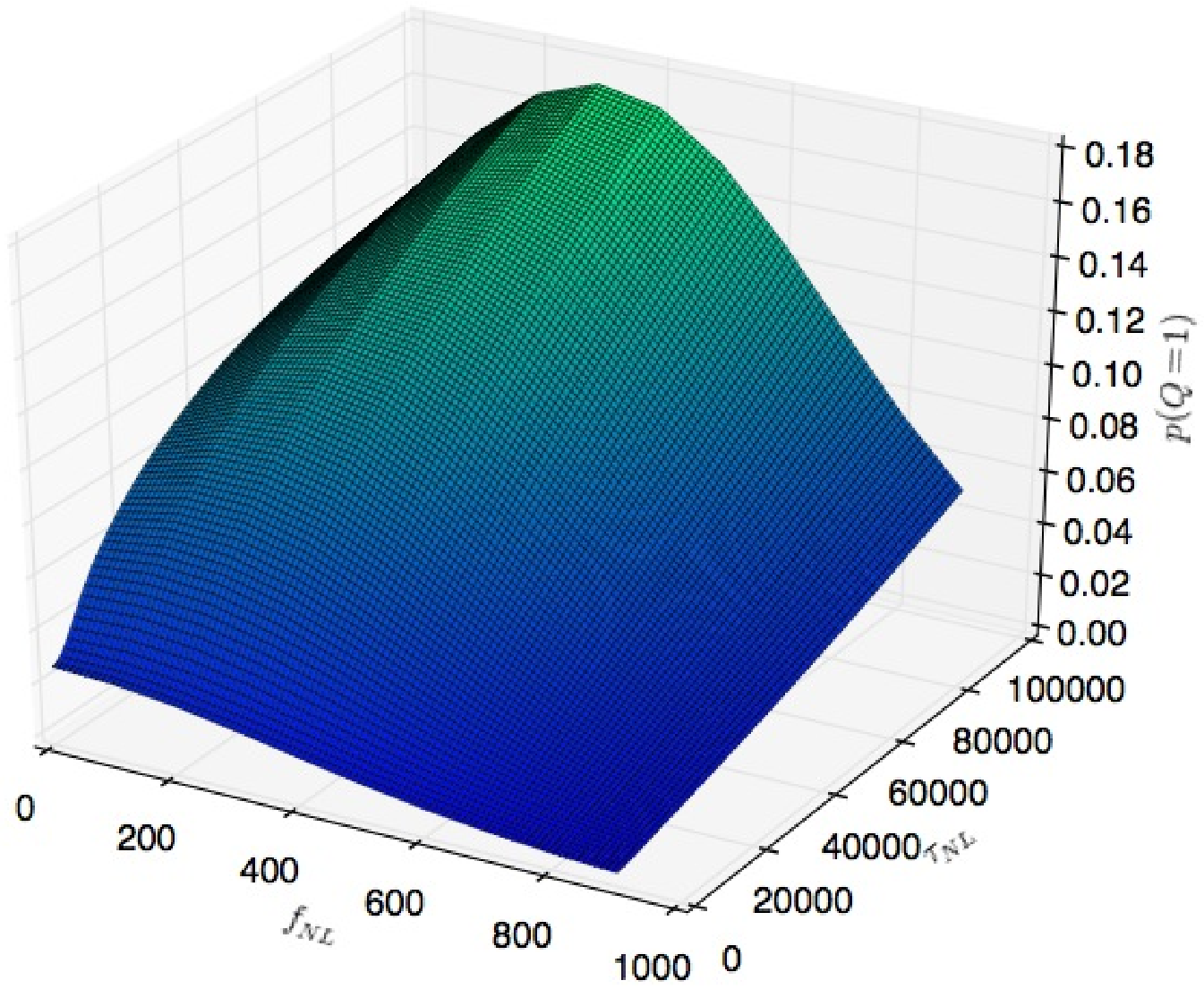}}
\end{center}
\caption{The probability distribution $p(Q)\dd Q$ of $Q = (6\fnl/5)^2/\tnl$ as a function of the non-Gaussianity parameters $\fnl$ and $\tnl$.}
\label{fig_pq_1}
\end{figure}

\begin{figure}
\begin{center}
\resizebox{0.5\hsize}{!}{\includegraphics{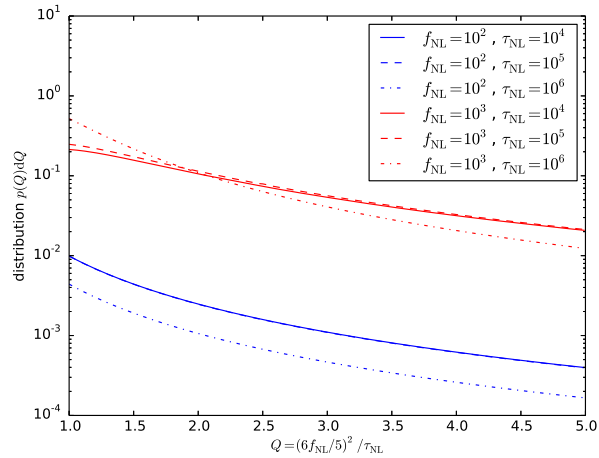}}
\end{center}
\caption{The probability distribution $p(Q)$ as a function of $Q$ for fixed non-Gaussianity parameter $\fnl = 10^2,~10^3$ (red and blue, respectively) and $\tnl = 10^4,~10^5,~10^6$ (solid, dashed and dash-dotted).}
\label{fig_pq_example}
\end{figure}

In Fig.~\ref{fig_pq_1} we are illustrating the ratio distribution as a function of $\fnl$ and $\tnl$ for $Q=1$, i.e. for the case where the SY-relation becomes an equality. The values for $\fnl$ run from 1 to $10^3$ and $\tnl$ runs from $1$ to $10^6$. The variances $\sigma_{\fnl}$ and $\sigma_{\tnl}$ are taken from the output of the Fisher matrix and correspond to $\sigma_{\fnl}=93$ and $\sigma_{\tnl}=7.5\times10^5$. We would like to point out the nice outcome, that the distribution has a clearly visible bumped line along the the SY-equality. Similarly, Fig.~\ref{fig_pq_example} shows a number of example distributions $p(Q)\dd Q$ for a choice of non-Gaussianity parameters $\fnl$ and $\tnl$. We let $Q$ run from 1 to 5 and fix the values $\fnl=10^2,~10^3$ and $\tnl=10^4,~10^5,~10^6$. 

\citet{Smidt:2010oq} study possible bounds on $A_\mathrm{NL}=1/Q$ based on a combination of CMB probes. The value of $\fnl=32$ suggested by WMAP7 \cite{Komatsu:2010fb} would imply that a a detection of $\tnl$ with Planck is possible if $Q<1/2$, and future experiments such as COrE \citep{2011arXiv1102.2181T} or EPIC \citep{2008arXiv0805.4207B} can probe regions of smaller trispectra, which might be relevant as a number of models predict small bi- and large trispectra, and could be a favourable for detecting non-Gaussianities. In our work we prefer to work with the probability distribution of $Q$ because for small values of $\fnl$ as suggested by Planck \citep{Planck-Collaboration:2013ve} one naturally obtains large values for $A=1/Q$.

\section{Bayesian evidence for a violation of the SY-equality}\label{sect_bayes}
\begin{figure}
\begin{center}
\resizebox{0.5\hsize}{!}{\includegraphics{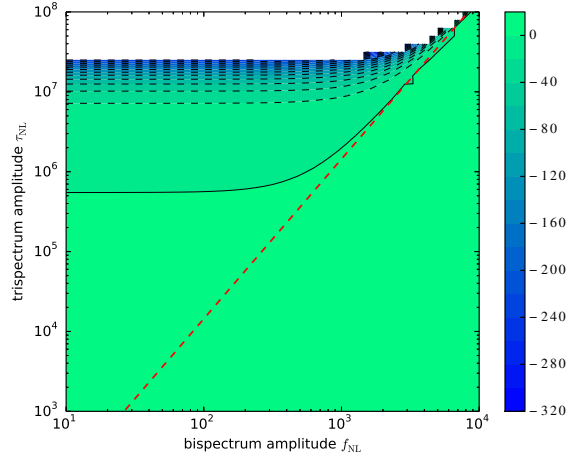}}
\end{center}
\caption{Logarithm of the Bayesian evidence ratio $E_=/E_\geq$, indicating that for most of the parameter range preference is given to the simpler hypothesis $E_=$, only in the parameter region $\tnl\gg\fnl$ the hypothesis $E_\geq$ is preferred.}
\label{fig_bayes}
\end{figure}

\spirou{An interesting quantity from a Bayesian point of view is the evidence ratio provided by a measurement comparing a model in which the SY-equality is fulfilled ($\tnl=(6\fnl/5)^2$) in contrast to the model with a SY-violation ($\tnl\geq (6\fnl/5)^2$). Following \citet{2007MNRAS.378...72T, 2008ConPh..49...71T} we define the evidences $E$ for either model,
\begin{eqnarray}
E_= & = & \int\dd\fnl^\prime\:p_=(\fnl^\prime)p_\mathrm{CMB}(\fnl^\prime)\\
E_\geq & = & \int\dd\fnl^\prime\:p_\geq(\fnl^\prime)p_\mathrm{CMB}(\fnl^\prime)
\end{eqnarray}
with a prior on the two non-Gaussianity parameters from the cosmic microwave background, whose functional shape we assume to be Gaussian. The two distributions $p_=(\fnl)\dd\fnl$ and $p_\geq(\fnl)\dd\fnl$ originate from a joint Gaussian on $\fnl$ and $\tnl$ with the Fisher-matrix as the inverse covariance where the conditions $\tnl=(6\fnl/5)^2$ and $\tnl\geq(6\fnl/5)^2$ are integrated out,
\begin{eqnarray}
p_=(\fnl^\prime) & = & \int\dd\tnl^\prime\:\mathcal{L}(\fnl-\fnl^\prime,\tnl-\tnl^\prime)\:\dirac\left(\tnl-(6/5\fnl)^2\right)\\
p_\geq(\fnl^\prime) & = & \int\dd\tnl^\prime\:\mathcal{L}(\fnl-\fnl^\prime,\tnl-\tnl^\prime)\:\heaviside\left(\tnl-(6/5\fnl)^2\right)
\end{eqnarray}
such that $E_\geq$ is equal to $\alpha$ up to the prior. Effectively, the SY-relation is used as a marginalisation condition. Finally, the Bayes ratio $B=E_=/E_\geq$ can be used to decide between the two models given the measurement and the prior, as it quantifies the model complexity needed for explaining the data. As a CMB-prior on $\fnl$, we assume a Gaussian with width $\sigma_{\fnl}\simeq 10$. Fig.~\ref{fig_bayes} suggests the preference of $E_=$ over $E_\geq$ over almost the entire parameter range, with the exception of $\tnl\gg\fnl$ in the upper left corner.}

\section{Summary}\label{sect_summary}
The topic of this paper is an investigation of inflationary bi- and trispectra by weak lensing, and testing of the SY-inequality relating the relative strengths of the inflationary bi- and trispectrum amplitudes using weak lensing as a mapping of the large-scale structure. Specifically, we consider the case of the projected Euclid weak lensing survey and choose a basic $w$CDM-cosmology as the background model.

\begin{enumerate}
\item{We compute weak lensing potential and weak lensing convergence spectra $C_\kappa$, bispectra $B_\kappa$ and trispectra $T_\kappa$ by Limber-projection from the CDM-polyspectra $P_\Phi$, $B_\Phi$ and $T_\Phi$ of the Newtonian gravitational potential $\Phi$. The non-Gaussianity model for the higher-order spectra are local non-Gaussianities parametrised with $\fnl$, $\gnl$ and $\tnl$. The weak lensing polyspectra reflect in their magnitude the perturbative ansatz by which they are generated and collect most of their amplitude at distances of $\sim1~\mathrm{Gpc}/h$, where the higher order polyspectra show a tendency to be generated at slightly smaller distances. Ratios of polyspectra where the transfer function has been divided out, nicely illustrate the reduction to products of spectra by application of the Wick theorem, as a pure power-law behaviour is recovered by this construction.}
\item{The signal to noise ratios $\Sigma_C$, $\Sigma_B$ and $\Sigma_T$ at which the polyspectra can be estimated with Euclid's weak lensing data are forecasted using a very efficient Monte-Carlo integration scheme for carrying out the configuration space summation. These integrations are carried out in flat polar coordinates with a Gaussian expression for the signal covariance. Whereas the first simplification should influence the result only weakly as most of the signal originates from sufficiently large multipoles, the second simplification has been shown to be violated in the investigation of dominating structure formation non-Gaussianities, but might be applicable in the case of weak inflationary non-Gaussianities and on low multipoles.}
\item{With a very similar integration scheme we compute a Fisher-matrix for the set of non-Gaussianity parameters $\fnl$, $\gnl$ and $\tnl$ such that a Gaussian likelihood $\mathcal{L}$ can be written down. Marginalisation over $\gnl$ yields the final likelihood $\mathcal{L}(\fnl,\tnl)$ which is the basis of the statistical investigations concerning the SY-inequality. The diagonal elements of the Fisher matrix are simply inverse squared signal to noise ratios due to the proportionality $B_\kappa\propto\fnl$ and $T_\kappa\propto\tnl$. For Gaussian covariances, the parameters $\fnl$ and $\tnl$ are statistically independent.}
\item{We quantify the degree of belief in the SY-relation with a set of inferred values for $\fnl$ and $\tnl$ and with statistical errors $\sigma_{\fnl}$ and $\sigma_{\tnl}$ by computing the Bayesian evidence that the SY-relation $\tnl\geq(6\fnl/5)^2$ is fulfilled. Euclid data would provide evidence in favour of the relation for $\tnl\gsim10^5$ and against the relation if $\fnl\gsim10^2$. For $\fnl<10^2$ and $\tnl\lsim10^5$ the Bayesian evidence is inconclusive and quite generally, larger non-Gaussianities allow for a better probing of the relation. \spirou{Comparing the Bayesian evidence of an equality in comparison to an inquality suggests that the equality is preferred as an explanation of the data given the amount of statistical error expected from the weak lensing measurement and that distinguishing between the two cases is difficult, except for extreme cases where $\tnl\gg\fnl$.}}
\item{We provide a computation of the probability that the quantity $Q\equiv(6\fnl/5)^2/\tnl$ is one, i.e. for an exact SY-relation. The distribution can be derived by generating a $\chi^2$-distribution for $\fnl^2$ and then by Mellin-transform for the ratio $\fnl^2/\tnl$. We observe, that the analytical probability distribution has a clearly visible bumped line along the SY-equality.}
\end{enumerate}

In summary, we would like to point out that constraining non-Gaussianities in weak lensing data is possible but the sensitivity is weaker compared to other probes. Nevertheless, for the small bispectrum parameter confirmed by Planck, $\tnl$ values of the order of $10^5$ would be needed to claim a satisfied SY-relation, and values smaller than that would not imply a violation, given the large experimental uncertainties. \spirou{If we assume that the non-linearity parameters are completely scale independent, then the Planck constraints of $-9.1< f_{\rm NL}<14$ and $\tau_{\rm NL}<2800$ (both bounds are quoted at the 95\% confidence limit) push us towards the region on the lower left hand side of Fig.~\ref{fig_sy_fnltnl}, where the observational data is not able to discriminate whether the Suyama-Yamaguchi inequality is saturated, holds or is broken. However if non-Gaussianity is larger on small scales, or if the sensitivity of weak lensing data can be significantly improved using tomography then a more positive conclusion might be reached.}

Despite the fact that we will not be able to see a violation of the inequality, if  $\tnl$ is large enough to be observed, then this together with the tight observational constraints on $\fnl$ will imply that the single-source relation is broken and instead $\tnl\gg\fnl^2$. Even though this is allowed by inflation, such a result would come as a surprise and be of great interest, since typically even multi-source scenarios predict a result which is close to the single-source equality, and a strong breaking is hard to realise for known models, e.g.~\cite{Peterson:2010mv,Elliston:2012wm,Leung:2013rza}, although examples can be constructed at the expense of fine tuning \citep{Ichikawa:2008ne, Byrnes:2008zy}.

\begin{figure}
\begin{center}
\resizebox{0.5\hsize}{!}{\includegraphics{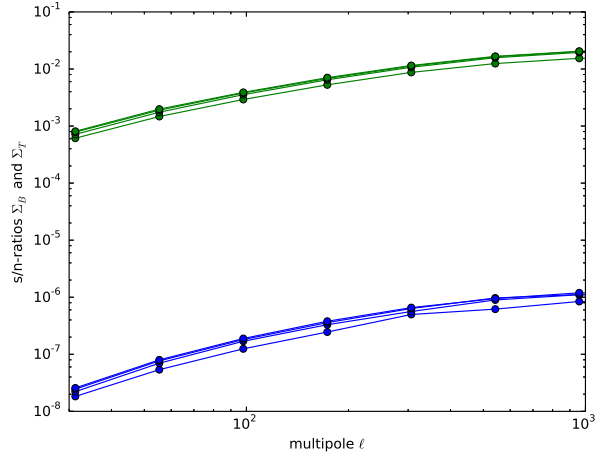}}
\end{center}
\caption{Cumulative signal to noise ratios $\Sigma_B$ (green lines) and $\Sigma_T$ (blue lines) for measuring the convergence bi- and trispectrum in a tomographic weak lensing survey, with $N=1,2,3,4$ (bottom to top) redshift bins.}
\label{fig_tomo}
\end{figure}

\spirou{As an outlook we provide a very coarse projection what levels of $\fnl$ and $\tnl$ can be probed by tomographic surveys \citep{1999ApJ...522L..21H,2004MNRAS.348..897T} with $N=2,3,4$ redshift bins which are chosen to contain equal fractions of the galaxy distribution, as a way of boosting the sensitivity, to decrease statistical errors and break degeneracies \citep{2008MNRAS.389..173K, Schaefer:2011dx}, in our case on the non-Gaussianity parameters. The binning was idealised with a fraction $1/n_\mathrm{bin}$ of galaxies in each of the $n_\mathrm{bin}$ bins, and without taking redshift-errors into account. The shape noise was assumed to be $n_\mathrm{bin}\times\sigma_\epsilon^2/n$ with the total number $n$ of galaxies per steradian and $\sigma\epsilon\simeq0.3$.} .Fig.~\ref{fig_tomo} shows the signal to noise ratio $\Sigma_B$ and $\Sigma_T$ for measuring local weak lensing bi- and trispectra, respectively, and at the same time those numbers correspond to the inverse statistical errors $\sigma_{\fnl}$ and $\sigma_{\tnl}$ because of the proportionality $B_\kappa\propto\fnl$ and $T_\kappa\propto\tnl$. \spirou{Taking the full covariance between lensing bi- and trispectra into account yields an improvement on the error on $\fnl$ by about 40\% and on $\tnl$ by about 50\%. These numbers are valid for the planned Euclid-survey.} Of course, many systematical effects become important, related to the measurement itself \citep{2011arXiv1105.1075S,2013arXiv1303.1808H}, to structure formation non-Gaussianities at low redshifts \citep[which can in principle be controlled with good priors on cosmological parameters, ][]{Schafer:2012kl}, or to the numerics of the polyspectrum estimation \citep{2011arXiv1104.0930S}.

\section*{Acknowledgements}
AG's and BMS's work was supported by the German Research Foundation (DFG) within the framework of the excellence initiative through GSFP$+$ at Heidelberg and the International Max Planck Research School for astronomy and cosmic physics. LH was supported by the Swiss Science Foundation and CTB acknowledges support from the Royal Society. We would like to thank Gero J{\"u}rgens for his support concerning the expressions for the covariance of bi- and trispectra. LH would like to thank to Claudia de Rham and Raquel Ribeiro for very useful discussions. Finally, we would like to express our gratitude to the anonymous referee for thoughtful questions.

\bibliography{bibtex/aamnem,bibtex/references}
\bibliographystyle{mn2e}

\appendix

\bsp

\label{lastpage}

\end{document}